\documentclass[twocolumn,superscriptaddress, prb, footinbib]{revtex4}

\usepackage{graphicx}
\usepackage{amsmath}
\usepackage{textcomp}
\usepackage{hyperref}
\usepackage{textcomp}
\usepackage{color}
\begin{document}

\title{First-principles evaluation of Multi-valent cation insertion into Orthorhombic V$_2$O$_5$}

\author{Gopalakrishnan Sai Gautam} \email{gautam91@mit.edu}
\affiliation{
Department of Materials Science and Engineering, Massachusetts Institute of Technology, Cambridge, MA 02139, USA}

\author{Pieremanuele Canepa}\affiliation{
Department of Materials Science and Engineering, Massachusetts Institute of Technology, Cambridge, MA 02139, USA}

\author{Rahul Malik}\affiliation{
Department of Materials Science and Engineering, Massachusetts Institute of Technology, Cambridge, MA 02139, USA}

\author{Miao Liu}\affiliation{
Environmental Energy Technologies Division, Lawrence Berkeley
National Laboratory, CA 94720, USA}

\author{Kristin Persson}\affiliation{
Environmental Energy Technologies Division, Lawrence Berkeley
National Laboratory, CA 94720, USA}

\author{Gerbrand Ceder} \email{gceder@berkeley.edu; gceder@lbl.gov}
\affiliation{
Department of Materials Science and Engineering, University of California Berkeley, CA 94720, USA}
\affiliation{
Materials Science Division, Lawrence Berkeley National Laboratory, CA 94720, USA}

\begin{abstract}
\textbf{
A systematic first principles evaluation of the insertion behavior of multi-valent cations in orthorhombic V$_2$O$_5$ is performed. Layer spacing, voltage, phase stability, and ion mobility are computed for Li$^+$, Mg$^{2+}$, Zn$^{2+}$, Ca$^{2+}$, and Al$^{3+}$ intercalation in the $\alpha$ and $\delta$ polymorphs.} 
\end{abstract}

\maketitle


\section*{}
\label{sec:Main}

A promising and realistic strategy to improve the energy density beyond the capability of current Li-ion battery technology is to transition to a battery architecture based on shuttling multi-valent (MV) ions (e.g. Mg$^{2+}$ or Ca$^{2+}$) between an intercalation cathode host and MV metal anode.\cite{Noorden2014,Aurbach2000a} Specifically, improvement in the volumetric energy density arises from the combination of using a multi-valent metal as the anode as opposed to an insertion structure (e.g. $\sim$~3833~mAh/cm$^{3}$ volumetric capacity for Mg metal compared to $\sim$~800~mAh/cm$^{3}$ for graphite), and storing more charge per ion in the cathode.\cite{Shterenberg2014a,Yoo2013}

One of the major bottlenecks preventing the development of MV battery technology, however, is the poor electrochemical performance of potential MV cathode materials, thought to originate predominantly from poor MV ion mobility in the intercalation host structure.\cite{Yoo2013,Gershinsky2013,Amatucci2001a} Moreover, the simultaneous challenge of developing functioning MV anodes and electrolytes compatible with candidate cathode materials has limited the ability to experimentally isolate and evaluate cathode electrochemical performance,\cite{Muldoon2012} and as such there is a general dearth of reliable data on MV ion intercalation in the literature to date to guide the ongoing search for new MV cathode materials with improved performance.

\begin{figure}[]
\centering
\includegraphics[width=\columnwidth]{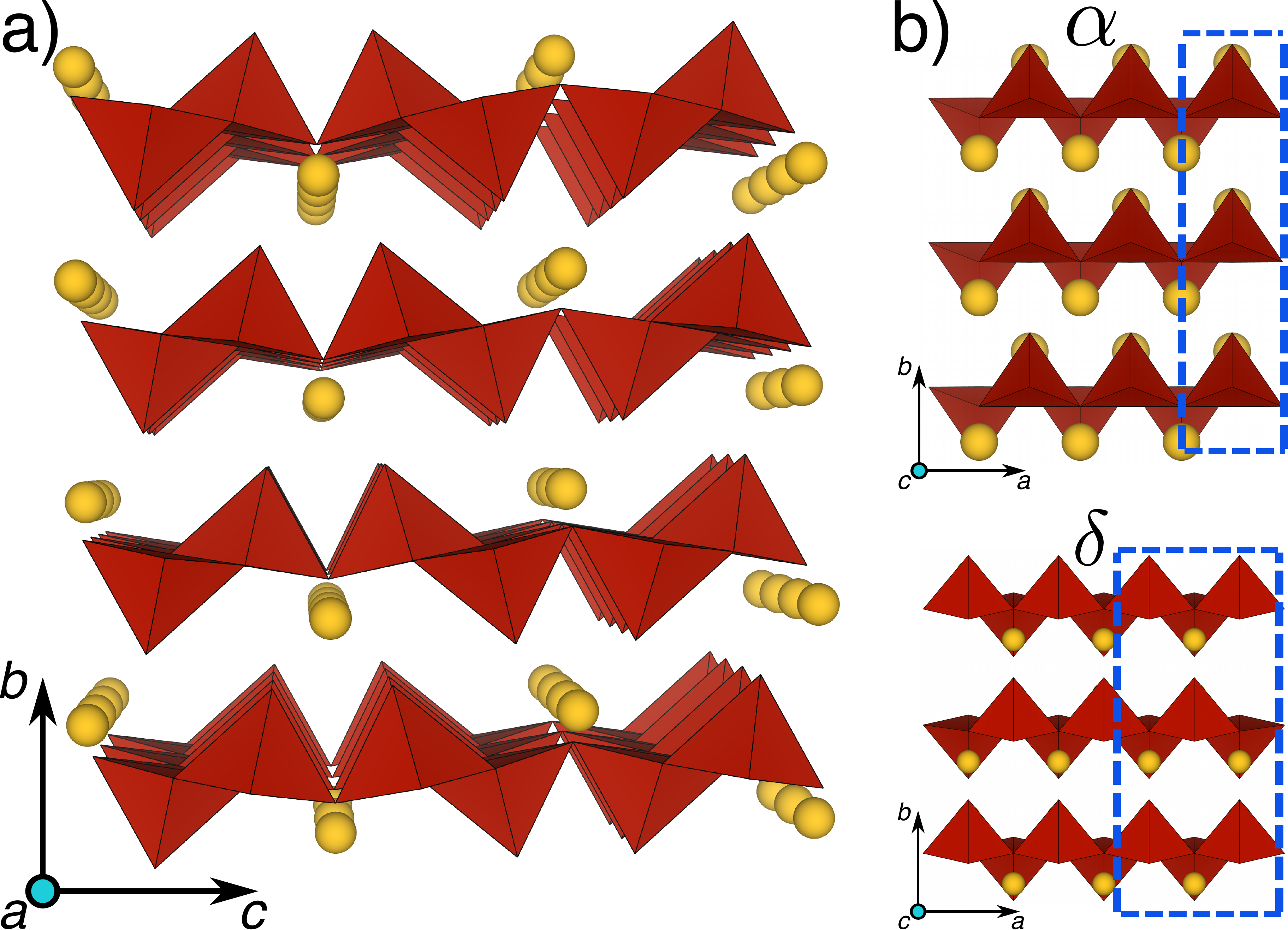}
\caption{
\label{fig:1}
\textbf{a) The V$_2$O$_5$ structure of both the $\alpha$ and $\delta$ polymorphs on the \textit{b}$-$\textit{c} plane with the yellow spheres indicating the intercalant sites while b) shows the $\alpha$ and $\delta$ polymorphs on the \textit{a}$-$\textit{b} plane. As indicated by the dashed blue regions, both the polymorphs differ by a change in the stacking of the V$_2$O$_5$ layers.}}
\end{figure}

Nevertheless, reversible electrochemical Mg$^{2+}$ intercalation has been successfully demonstrated in a handful of cathode hosts, namely Chevrel Mo$_6$S$_8$ ($\sim$~135~mAh/g capacity at $\sim$~1.0~$-$~1.3~V vs. Mg metal),\cite{Aurbach2000a} as well as layered V$_2$O$_5$ ($\sim$~150~mAh/g at $\sim$~2.3~$-$~2.6~V)\cite{Gershinsky2013,Amatucci2001a} and MoO$_3$ ($\sim$~220~mAh/g at $\sim$~1.7~$-$~2.8~V).\cite{Gershinsky2013} The orthorhombic V$_2$O$_5$ structure is especially interesting because it has also demonstrated the ability to reversibly intercalate Ca$^{2+}$ and Y$^{3+}$ in addition to Mg$^{2+}$ ions.\cite{Amatucci2001a} First principles calculations (described in more detail in the supplementary information\footnote{Electronic Supplementary Information available online: http://dx.doi.org/10.1039/C5CC04947D}) have proven to be an accurate and effective method to systematically assess the electrochemical properties of Li-ion batteries,\cite{Kang2006,VanDerVen2013,Meng2013} and have also been used to study the process of ion intercalation in layered materials, such as graphite{\cite{Yoon2015}} and V$_2$O$_5$.\cite{SaiGautam2015,Wang2013,Zhou2014b,Carrasco2014} In this work, we have performed a systematic first principles study of MV ion intercalation in the orthorhombic $\alpha-$ and $\delta-$V$_2$O$_5$ polymorphs by evaluating the structural change, voltage, thermodynamic stability, and intercalant mobility for Li$^{+}$, Mg$^{2+}$, Zn$^{2+}$, Ca$^{2+}$, and Al$^{3+}$ insertion and comparing to data in the literature when available.

The crystal structure and intercalation sites of the $\alpha-$ and $\delta-$V$_2$O$_5$ polymorphs\cite{Delmas1994,Millet1998,Enjalbert1986,Bachmann1961} are shown in Fig.~\ref{fig:1}. Perpendicular to the \textit{b}-axis (i.e. in the \textit{a}$-$\textit{c} plane), the orthorhombic V$_2$O$_5$ structure consists of layers of alternating corner$-$ and edge$-$sharing VO$_5$ pyramids (shown in red), each consisting of 4 V$-$O bonds that form the base and one short V$=$O bond that forms the apex. The intercalation sites (yellow spheres) are situated in between the layers, and assuming no limitation in the number of redox centers, the theoretical gravimetric capacities for \textit{A}V$_2$O$_5$ where \textit{A}~$=$~Li, Mg, Zn, Ca and Al are 142, 260, 217, 242 and 385~mAh/g, respectively. Structurally, the main difference between the $\alpha$ and $\delta$ polymorphs is a shift in the layer stacking, indicated by the dashed blue lines in Fig.~\ref{fig:1}b, with alternate V$_2$O$_5$ layers displaced in the \textit{a}-direction by half a lattice spacing, accompanied by a change in the interlayer distance and the anion coordination environment of the intercalation sites.\cite{Delmas1994} While 8 oxygen atoms coordinate the intercalant ion in $\alpha$ (for Mg, there are two Mg$-$O bonds with length $\sim$~2.11~$\textrm{\AA}$, two with $\sim$~2.39~$\textrm{\AA}$, and four with $\sim$~2.46~$\textrm{\AA}$, respectively), ``4+2'' oxygen atoms coordinate the intercalant in $\delta$ (for Mg, there are four Mg-O bonds with length $\sim$~2.05~$-$~2.07~$\textrm{\AA}$, and two with $\sim$~2.33~$\textrm{\AA}$).

\begin{figure}[]
\includegraphics[width=\columnwidth]{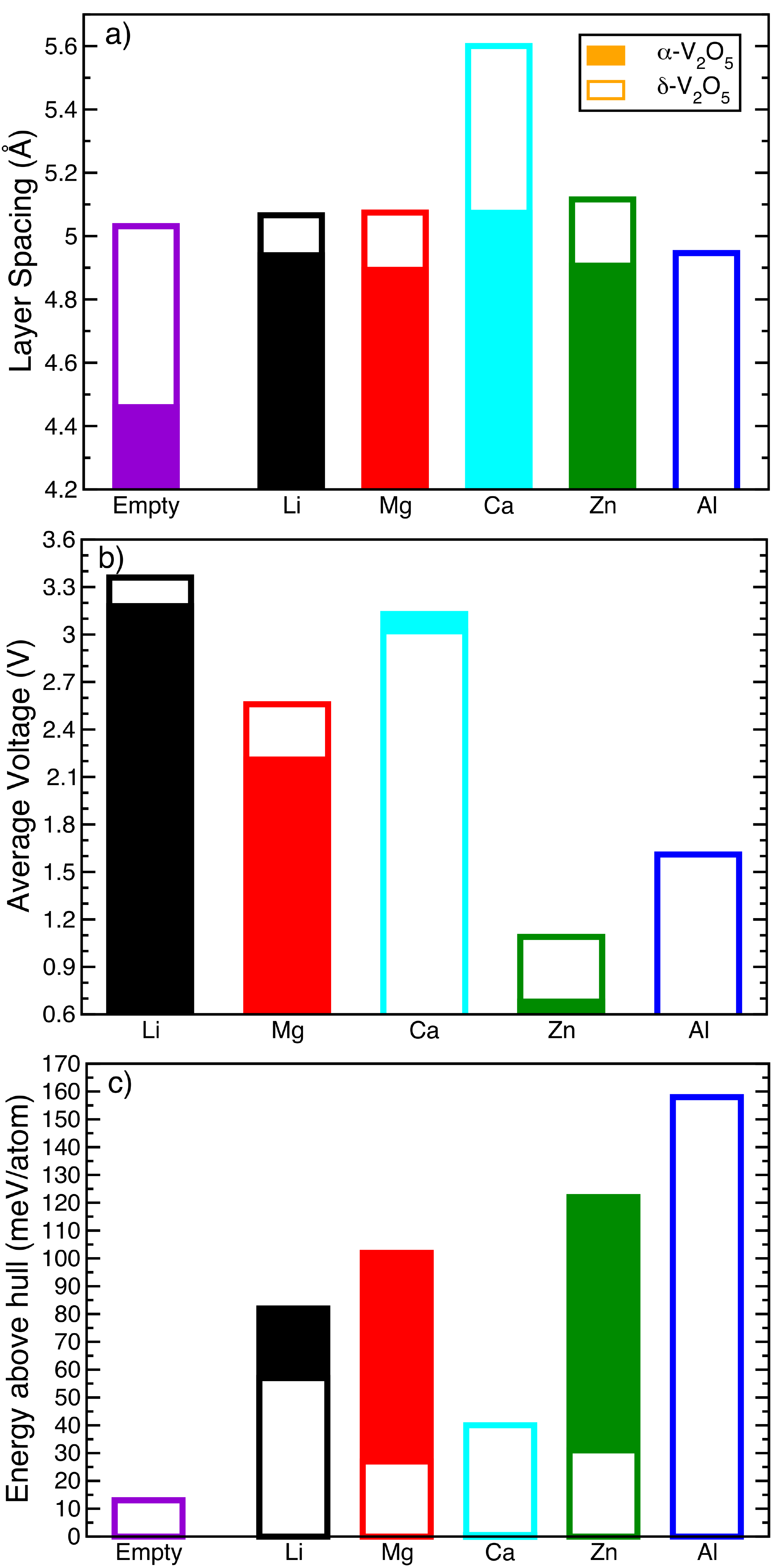}
\caption{ 
\label{fig:2}
 \textbf{a) Plots the layer spacing values for the empty and intercalated versions of \textit{A}V$_2$O$_5$ (\textit{A}~$=$~Li, Mg, Ca, Zn and Al) for both the $\alpha$ and $\delta$ polymorphs. b) Displays the calculated average voltage values for the intercalation of the different ions and c) shows the energy above hull, which quantifies the stability of a structure, for the empty and intercalated versions of $\alpha$ and $\delta$. The filled regions in all the graphs correspond to the $\alpha$ structure while the hollow regions correspond to the $\delta$ structure. Note that the energy above hull for $\alpha-$CaV$_2$O$_5$ is 0~meV/atom, implying that it is a ground state configuration in the Ca-V-O system.}}
\end{figure}

In Fig.~\ref{fig:2}a, the interlayer spacings in the $\alpha$ and $\delta$ polymorphs (filled and hollow bars, respectively) are shown for empty V$_2$O$_5$ and intercalated \textit{A}V$_2$O$_5$, where \textit{A}~$=$~Li, Mg, Zn, Ca, and Al. To better capture the increased effect of van der Waals effects in the deintercalated limit, the interlayer spacings for empty V$_2$O$_5$ (4.46~$\textrm{\AA}$ for $\alpha$; 5.03~$\textrm{\AA}$ for $\delta$) are calculated using the vdW-DF2 functional\cite{Lee2010,Klime2011} rather than standard DFT as the latter significantly overestimates this spacing (4.75~$\textrm{\AA}$ for $\alpha$; 5.27~$\textrm{\AA}$ for $\delta$) compared to experiment (4.37~$\textrm{\AA}$ for $\alpha$).\cite{SaiGautam2015,Carrasco2014,Enjalbert1986} As detailed in the supplementary information, Al$^{3+}$ intercalation in the $\alpha-$V$_2$O$_5$ structure is found to be mechanically unstable and relaxes to the $\delta$ polymorph in our calculations, and we therefore remove it from further consideration in this study.

At the same intercalant composition, the $\delta$ structures consistently have larger layer spacings than $\alpha$, $\sim$~3~$-$~5~\% larger for Li, Mg, and Zn and $\sim$~10~$-$~12~\% for Ca and empty V$_2$O$_5$. With the exception of Ca intercalation, which increases the layer spacing by more than 10~\% in both polymorphs, the change in the layer spacing is much smaller in $\delta$ than $\alpha$, less than 2~\% for Li$^{+}$, Mg$^{2+}$, Zn$^{2+}$, and Al$^{3+}$ intercalation in $\delta-$V$_2$O$_5$ compared to $\sim$~9~$-$~14~\% for Li$^{+}$, Mg$^{2+}$, Zn$^{2+}$, and Ca$^{2+}$ in $\alpha-$V$_2$O$_5$. The behavior for Ca$^{2+}$ is consistent with intercalation in the spinel system,\cite{Liu2014a} where the volume change is also much larger than for Li$^{+}$, Mg$^{2+}$, Zn$^{2+}$, and Al$^{3+}$ intercalation, and in general may be attributed to the larger ionic radius of Ca$^{2+}$ in comparison to the other ions.\cite{Shannon1969} Al$^{3+}$ intercalation in $\delta-$V$_2$O$_5$, in contrast to the other ions considered, is accompanied by a contraction of the layers, which is consistent with its small ionic radius and higher positive charge density that strengthens the attraction with nearby oxygen ions.

The average voltages of the compounds computed using the method of Aydinol \emph{et al}.\cite{Aydinol1997} are plotted in Fig.~\ref{fig:2}b and are referenced to the potential of the bulk metal of the corresponding intercalating ion (i.e., Li metal for Li$^{+}$ intercalation, etc.). The average voltages computed for Li, Mg, and Ca intercalation compare very well to available experimental data: $\sim$~3.2~$-$~3.4~V for Li measured by Delmas \emph{et al}.,\cite{Delmas1994} $\sim$~2.2~$-$~2.4~V for Mg measured by Gershinsky \emph{et al}.,\cite{Gershinsky2013} and $\sim$~2.4~$-$~3.1~V for Ca measured by Amatucci \emph{et al}.\cite{Amatucci2001a} In general, the Li polymorphs have the highest voltage, followed by Ca, Mg, Al, and Zn, which reflects both the same order and approximately the same potential difference indicated by the electrochemical series ($-$3.04~V vs. SHE for Li, $-$2.86~V for Ca, $-$2.37~V for Mg, $-$1.66~V for Al, and $-$0.76~V for Zn). In comparison, the voltage difference between the V$_2$O$_5$ polymorphs is much smaller for a given intercalation chemistry. For Li, Mg, and Zn the insertion voltage is higher in $\delta$ (3.36~V, 2.56~V, and 1.09~V, respectively) than in $\alpha$ (3.18~V, 2.21~V, and 0.68~V), unlike for Ca where $\alpha$ is higher (3.13~V for $\alpha$; 3.02~V for $\delta$).

Fig.~\ref{fig:2}c displays the energy above the convex ground state energy hull (E${^\wedge}$hull) of the deintercalated and intercalated V$_2$O$_5$ polymorphs with respect to the intercalant-V-O ternary phase diagram. The ternary ground state hulls were determined from the available calculated compounds in the Materials Project database.\cite{Jain2013} A predicted thermodynamically stable structure will have a E$^{\wedge}$hull value of 0~meV/atom while higher (more positive) E$^{\wedge}$hull values indicate greater instability, which may be reflected in experimental difficulties in synthesis or decomposition during battery operation. Note that the E$^{\wedge}$hull values calculated here reflect the ground state (i.e. 0~K), and entropy contributions, which scale with $k_\text{B}T$, can stabilize certain structures at higher temperatures.

In the deintercalated limit, V$_2$O$_5$ is thermodynamically stable in the $\alpha$ phase, but $\delta$ is only $\sim$~13~meV/atom higher in energy, indicating the possibility of metastability at room temperature. For Li intercalation, the $\alpha$ and $\delta$ structures are 82~meV/atom and 57~meV/atom more unstable than the ground state orthorhombic $\gamma-$LiV$_2$O$_5$ structure, which has a different orientation of the VO$_5$ pyramids\cite{Delmas1994} along the \textit{c}-direction shown in Fig~\ref{fig:1}a, but the $\delta$ structure can remain metastable and has shown to be reversibly cycled electrochemically.\cite{Delmas1994} $\delta-$MgV$_2$O$_5$, which has been synthesized experimentally,\cite{Millet1998} is only $\sim$~27~meV/atom more unstable (compared to $\sim$~102~meV/atom for $\alpha$) than the thermodynamic ground state, a two-phase equilibrium consisting of MgVO$_3$ and VO$_2$. Similarly $\delta$-ZnV$_2$O$_5$ is only $\sim$~31~meV/atom more unstable than the ground state (ZnO and VO$_2$), indicating that a metastable synthesis comparable to the Mg system may be possible. As Al intercalated $\alpha$-V$_2$O$_5$ displays mechanical instability in our calculations, when relaxed its energy is not defined, but the Al intercalated $\delta$-phase is $\sim$~158~meV/atom unstable compared to the ground state ternary equilibrium of Al$_2$O$_3$, VO$_2$ and V$_3$O$_5$. With the exception of $\alpha-$CaV$_2$O$_5$, which is the ground state in the intercalated Ca-V$_2$O$_5$ system, the $\delta$ structures tend to be more stable than $\alpha$ in the discharged state (by 25~meV/atom for Li; 75~meV/ atom for Mg; and 91~meV/atom for Zn), and accordingly the insertion voltages for $\delta$ are higher than $\alpha$ for Li, Mg, and Zn insertion but lower for Ca insertion, as observed in Fig~\ref{fig:2}b. Given that the intercalant sites in $\alpha$ and $\delta$ are coordinated by 8 and ``4+2'' oxygen atoms respectively, the stability of the discharged $\delta$-V$_2$O$_5$ structures for Li, Mg and Zn, and $\alpha$-V$_2$O$_5$ for Ca align well with the preferred coordination environment of the respective ions, as tabulated by Brown.\cite{Brown:tr0200} Hence for intercalant ions that prefer a lower coordination number (i.e., coordinated by a maximum of 6 neighboring atoms), an $\alpha \rightarrow \delta$ transition upon insertion in V$_2$O$_5$ is likely.

\begin{figure}[]
\includegraphics[width=\columnwidth]{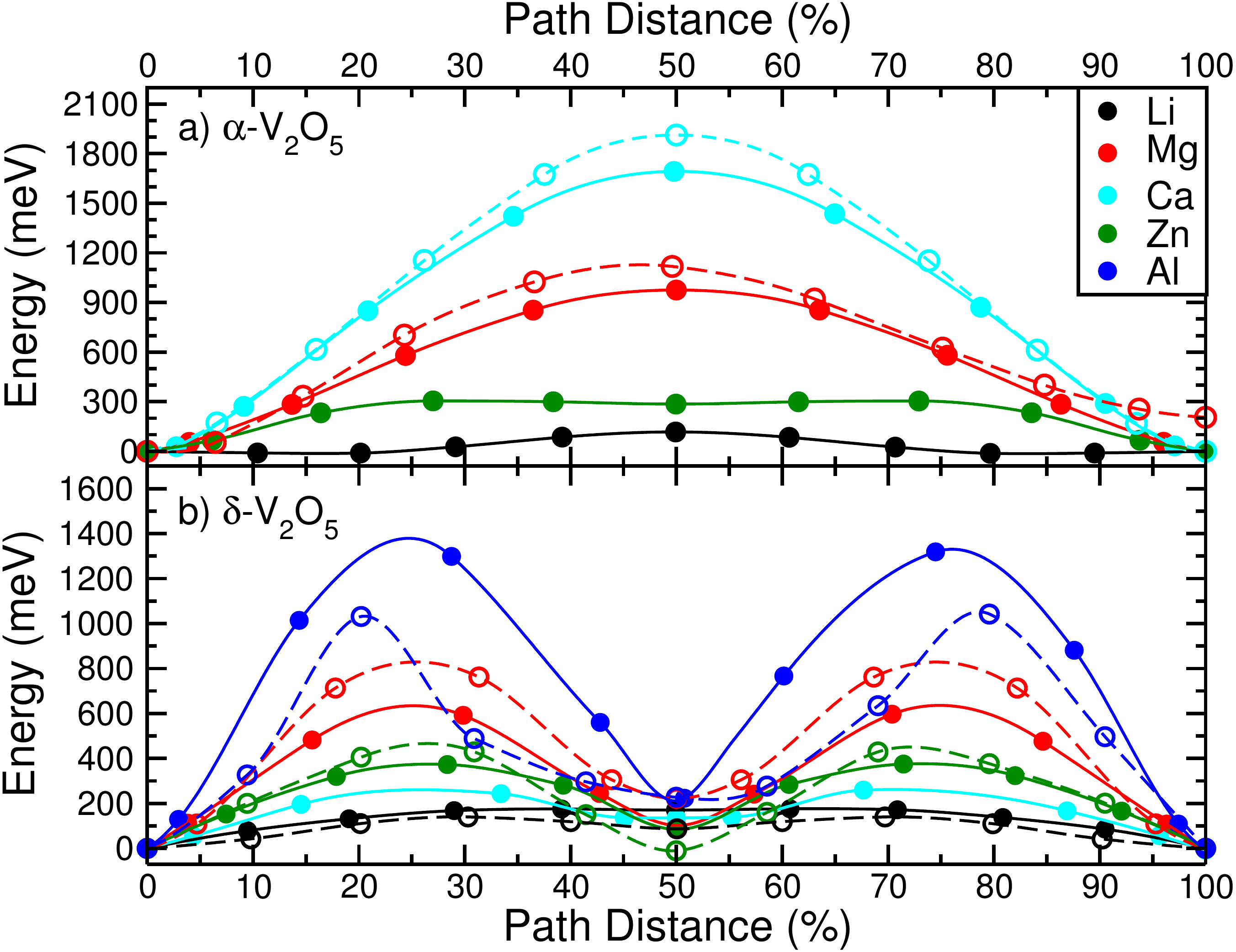}
\caption{ 
\label{fig:3}
\textbf{The activation barriers for the diffusion of the different intercalating ions in the $\alpha$ and $\delta$ polymorphs are plotted in a) and b) respectively. The solid lines correspond to the empty lattice limit (charged state) while the hollow lines correspond to the full lattice limit (discharged state).}}
\end{figure}

Fig.~\ref{fig:3} displays the migration energies for intercalant diffusion along the \textit{a}-direction in the $\alpha$ (Fig.~\ref{fig:3}a) and $\delta$ (Fig.~\ref{fig:3}b) polymorphs plotted against the normalized path distance calculated with the Nudged Elastic Band method.\cite{Sheppard2008} The solid lines correspond to migration energies obtained in the empty lattice limit (charged state), and the dashed lines correspond to the fully intercalated limit (discharged state). As elaborated upon in the supplementary information, converging the migration energies in structures that exhibit a high degree of thermodynamic instability may not be possible, as was the case for Li, Mg, and Zn in the intercalated $\alpha-$V$_2$O$_5$ structure, and for Ca in the intercalated $\delta-$V$_2$O$_5$ structure. In lieu of determining the Mg migration barrier in the fully discharged $\alpha-$V$_2$O$_5$ structure, we have computed the energy for Mg migration in a half intercalated structure with a specific ordering of Mg ions, referred to as the ``$\epsilon$'' phase, which has also been observed in the Li-V$_2$O$_5$ system.\cite{Delmas1994}

In Fig~\ref{fig:3}, the maximum energy difference encountered along the diffusion path defines the migration barrier (E$_\text{m}$), which provides an approximate estimate of the ionic diffusivity. As a guide, at room temperature, E$_\text{m}\sim$~525~meV corresponds to a diffusivity of $\sim$~10$^{-12}$~cm$^2$s$^{-1}$, and a 60~meV increase (decrease) in the migration energy corresponds to an order of magnitude decrease (increase) in diffusivity. Due to stronger interactions between a multivalent intercalant and the surrounding anion environment, the migration barriers within the same host structure, for example Al$^{3+}$, are generally higher than the divalent ion barriers (Mg$^{2+}$, Zn$^{2+}$, Ca$^{2+}$), which are generally higher than the barriers for Li$^{+}$. For the divalent intercalants, the trend in the migration barriers is Ca$^{2+}$ ($\sim$~1700$-$1900~meV)~$>$~Mg$^{2+}$ ($\sim$~975$-$1100~meV)~$>$~Zn$^{2+}$ ($\sim$~305~meV) in the $\alpha$-phase, but Mg$^{2+}$ ($\sim$~600$-$800~meV)~$>$~Zn$^{2+}$ ($\sim$~375$-$425~meV)~$>$~Ca$^{2+}$ ($\sim$~200~meV) in the $\delta$ phase. The energy above the hull (Fig~\ref{fig:2}c) ranked from the lowest to highest reflects this same trend, with Ca$^{2+}$~$>$~Mg$^{2+}$~$>$~Zn$^{2+}$ in $\alpha$ and Mg$^{2+}$~$>$~Zn$^{2+}$~$>$~Ca$^{2+}$ for $\delta$, and highlights the positive correlation between high intercalant mobility and low thermodynamic stability. For both V$_2$O$_5$ polymorphs considered, the change in the migration barrier from the deintercalated to intercalated limit for the same diffusing species is much smaller than the variation across intercalating ions.

Although the $\alpha$ and $\delta$ polymorphs of V$_2$O$_5$ are structurally very similar as earlier discussed, the anion coordination environment and therefore diffusion topology of the migrating intercalant vary significantly, which accounts for the different shape of the migration energies seen in Fig~\ref{fig:3}a and Fig~\ref{fig:3}b. In the $\alpha$ phase, the stable insertion site is coordinated by 8 oxygen anions which is connected to the adjacent insertion site along the \textit{a}-axis by a 3-coordinated shared face. The shape of the migration energies shown in Fig.~\ref{fig:3}a, therefore, reflect the change in coordination of 8$\rightarrow$3$\rightarrow$8 encountered by the diffusing species with the migration barrier corresponding to passing through the shared face. For the $\delta$ phase, the stable insertion site adopts a ``4+2'' coordination and shares a corner with the adjacent insertion site along the \textit{a}-axis. To migrate to this site, the intercalant passes through a 3-coordinated face shared with an intermediate 5-coordinated (pyramidal) site, and finally performs a symmetric hop to the next insertion site. The change in the anion coordination along the diffusion path is then ``4+2''$\rightarrow$3$\rightarrow$5$\rightarrow$3$\rightarrow$``4+2'', where occupation of the intermediate pyramidal site corresponds to a local minimum in the migration energy, as is reflected in Fig~\ref{fig:3}b. Overall, the migration barriers are also lower in the $\delta$ phase compared to $\alpha$ (significantly lower for some cases), which we attribute in large part to the smaller coordination change during the migration process encountered in $\delta$. Also, the change in the relative order of the migration barriers of divalent ions between $\alpha$ (Ca~$>$~Mg, Zn) and $\delta$ (Mg, Zn~$>$~Ca) can be explained by the correlation between the ``preferred'' coordination environments of the respective ions and the available anion coordination environments around the intercalation sites.\cite{Rong2015} In a given structure, migration barriers are higher for an ion whose preferred coordination aligns with that of the coordination environment available for the intercalant site compared to an ion whose preferred coordination is different from that present in the structure. For example, Ca is in its preferred 8-coordinated site in $\alpha$ and hence has higher barriers than Mg and Zn, which are not in their respectively preferred 6 and 4 coordinated sites. Whereas in $\delta$, Ca is present in an unfavored ``4+2'' coordinated site and hence has lower barriers than either of Mg or Zn, which are closer to their preferred coordination environments. Our results thus lend support to the hypothesis that coordination of the intercalation site is a good screening criterion for identifying fast multi-valent cation diffusers. 

An ideal MV cathode intercalation host must possess several properties$-$high capacity, high insertion voltage, and MV ion mobility, while simultaneously minimal structural change and thermodynamic instability. From the systematic first-principles study performed in this work, we are able to evaluate all of the candidate materials across each of these criteria. On the basis of ion mobility, Al$^{3+}$ intercalation appears unfeasible at room temperature in V$_2$O$_5$ due to its prohibitively high migration barriers, and although Zn$^{2+}$ intercalation is determined to be facile in both polymorphs and relatively stable in the $\delta$ phase, the insertion voltage is low. Mobility of Mg$^{2+}$ and Ca$^{2+}$ is determined to be poor in the $\alpha$ phase, but intercalation of these ions in the $\delta$ phase appear most promising, with sufficiently high voltage (3.02~V for Ca, and 2.56~V for Mg) and mobility (E$_\text{m}\sim$~200~meV for Ca and $\sim$~600$-$800~meV for Mg) albeit with moderate thermodynamic instability (27~meV/atom for Mg and 40~meV/atom for Ca above the ground state hull in the discharged state).

\begin{acknowledgements}
The current work is fully supported by the Joint Center for Energy Storage Research (JCESR), an Energy Innovation Hub funded by the U.S. Department of Energy, Office of Science and Basic Energy Sciences. This study is supported by Subcontract 3F-31144. The authors would like to thank the National Energy Research Scientific Computing Center (NERSC) for providing computing resources.
\end{acknowledgements}


\bibliography{library} 

\end{document}